\documentclass[aps,manuscript,showpacs]{revtex4}

\usepackage{graphicx}
\usepackage{epsfig}
\usepackage{amssymb}
\usepackage{amsmath}
\usepackage{psfrag}

\newcommand{\rg}{{\bf r}}
\newcommand{\Eg}{{\bf E}}
\newcommand{\Gg}{{\bf G}}

\newcommand{\Mass}{{\bf \Sigma}}
\newcommand{\Trans}{{\bf \Delta}}
\newcommand{\Ig}{{\bf I}}

\newcommand{\kg}{{\bf k}}
\newcommand{\Rg}{{\bf R}}
\newcommand{\dd}{{\mathrm{d}}}

\newcommand{\epseff}{\epsilon_{\mathrm{eff}}}
\newcommand{\neff}{n_{\mathrm{eff}}}
\newcommand{\keff}{k_{\mathrm{eff}}}
\newcommand{\lext}{\ell_{\mathrm{ext}}}
\newcommand{\labs}{\ell_{\mathrm{abs}}}
\newcommand{\ls}{\ell_{\mathrm{s}}}
\newcommand{\Imag}{\mathrm{Im} \, }
\newcommand{\Real}{\mathrm{Re} \, }

\begin{document}

\title{On the density of states and extinction mean free path of waves in random media: Dispersion relations and sum rules}
\author{R. Carminati$^1$}
\email{remi.carminati@espci.fr}
\author{M. Donaire$^2$}
\author{J.J. S\'aenz$^{2,3}$}
\affiliation{$^{1}$Laboratoire d'Optique Physique, ESPCI, CNRS, 10 rue Vauquelin,
75231 Paris Cedex 05, France\\
$^{2}$Departamento de F\'\i sica de la Materia Condensada and Instituto
``Nicol\'as Cabrera'', Universidad Aut\'onoma de Madrid, 28049 Madrid,
Spain\\
$^{3}$Donostia International Physics Center (DIPC),
 20018 Donostia-San Sebastian, Spain}

\begin{abstract}
We establish a fundamental relationship between the averaged density of states and the extinction mean free path of wave propagating in random media. From the principle of causality and the Kramers-Kronig relations, we show that both quantities are connected by dispersion relations and are constrained by a frequency sum rule. The results are valid under very general conditions and should be helpful in the analysis of measurements of wave transport through complex systems and in the design of randomly or periodically structured materials with specific transport properties.
\end{abstract}


\maketitle

%INTRODUCTION

Fundamental questions in coherent transport of electromagnetic, electronic or
acoustic waves~\cite{ShengBook,AkkermansBook} as well as applications to
imaging in complex media~\cite{SebbahBook} have made wave propagation in random
media a central issue in physics. Randomly or periodically structured materials
allow to design media or devices with unconventional properties. This includes
photonic crystals~\cite{WinnBook}, metamaterials for
electromagnetic~\cite{LakhtakiaBook,Smith04} or acoustic waves~\cite{Sheng00},
strongly correlated disordered systems~\cite{Juanjo04,Lopez07}, or materials
generating non-diffusive transport~\cite{Wiersma08}.

The density of states (DOS) and the extinction mean free path (MFP) are fundamental concepts in coherent wave transport. The DOS shapes many macroscopic transport properties~\cite{Lagen96}. The local density of states (LDOS) drives the spontaneous emission of light~\cite{Sipe84}, and is at the root of novel imaging techniques base on field correlations~\cite{Weaver01}. Fluctuations in the DOS or LDOS characterize the transport regime~\cite{Been94}, speckle patterns~\cite{Skipetrov06} or the local structure of a complex medium~\cite{Luis07}. The extinction MFP $\lext$, defined by $\lext^{-1} =\ls^{-1} +\labs^{-1}$ with $\ls$ and $\labs$ the scattering and absorption mean free paths, describes the attenuation of the averaged (or coherent) field. When absorption is negligible, the extinction MFP equals the scattering MFP. The latter is an important quantity since $k\ls$ is a measure of the strength of scattering, $k$ being the wavenumber in the medium. The spectral behavior of $\ls$ and the DOS were put forward in early studies of Anderson localization of light~\cite{John}. Moreover, the transition to localization in three dimensions is expected when $k\ls \lesssim 1$, according to the Ioffe-Regel criterion~\cite{ShengBook}.

In this Letter, we establish a fundamental relationship between the averaged DOS in a random medium and the extinction MFP. As a consequence of causality and the Kramers-Kronig relations, we show that both quantities are connected by dispersion relations and are constrained by a frequency sum rule. We focus the derivation on light propagation in scattering media, but the results are valid under very general conditions and should be applicable to any kind of waves.

%EFFECTIVE INDEX

Consider a scattering medium made of scatterers randomly distributed in free space (or in an otherwise homogeneous background medium).  The (dyadic) Green function (or electric-field susceptibility) describes the response at point $\rg$, and at a given frequency $\omega$, to a point electric-dipole source ${\bf p}$ located at point $\rg^\prime$ through the relation $\Eg(\rg) = \mu_0 \omega^2 \, \Gg(\rg,\rg^\prime) {\bf p}$. In free-space, the Green function reads~:
\begin{equation}
\Gg_0(\Rg) = PV \left [ \Ig+\frac{1}{k_0^2}\nabla\nabla \right ] \, \frac{\exp(ik_0R)}{4\pi R} - \frac{\Ig}{3 k_0^2} \delta(\Rg)
\label{eq:G0}
\end{equation}
where $\Ig$ is the unit tensor, $k_0=\omega/c$, with $c$ the speed of light in vacuum (or in the homogeneous background medium), $\Rg=\rg-\rg^\prime$, $R=|\Rg|$ and $PV$ stands for principal value. Its Fourier transform is given by $\Gg_0(\kg) =[(k^2-k_0^2) \Ig - \kg\kg]^{-1}$~\cite{Lagen96}. In a random medium, after averaging over the positions of the scatterers and assuming statistical translational invariance, the averaged Green function $\langle \Gg(\rg-\rg^\prime) \rangle$ obeys the Dyson equation~\cite{ShengBook,Frisch}~:
\begin{equation}
\langle \Gg(\kg) \rangle = \Gg_0(\kg) + \Gg_0(\kg) \, {\bf \Mass}(\kg) \, \langle \Gg(\kg) \rangle
\label{eq:Dyson}
\end{equation}
in which $\Mass(\kg)$ is the self energy (or mass operator) containing the sum of all multiply connected scattering events~\cite{Frisch}.
From Eq.~(\ref{eq:Dyson}) and the expression of $\Gg_0(\kg)$, the averaged Green function can be written~:
\begin{eqnarray}
 \langle \Gg(\kg) \rangle &=& \frac{\Ig}{(k^2 -k_0^2) \Ig -\kg\kg - \Mass(\kg)} \nonumber \\
 &=&  \frac{\Trans(\kg)}{k^2 \Ig -k_0^2 \, \boldsymbol{\epsilon}^\perp_{\mathrm{eff}}(\kg)}  -\frac{\kg\kg}{k_0^2 \, k^2 \, \boldsymbol{\epsilon}^\parallel_{\mathrm{eff}}(\kg)}
\end{eqnarray}
where $\Trans(\kg) = \Ig-\kg\kg/k^2$ is the transverse projection operator.
In the last equality, we have identified the effective dielectric function~:
\begin{equation}
\boldsymbol{\epsilon}_{\mathrm{eff}}(\kg) =  \Ig + \frac{\Mass(\kg)}{k_0^2}
\end{equation}
and its transverse and longitudinal projections
$\boldsymbol{\epsilon}^\perp_{\mathrm{eff}}(\kg)$ and
$\boldsymbol{\epsilon}^\parallel_{\mathrm{eff}}(\kg)$. This quantity drives the
propagation of the averaged (or coherent) field in the random medium, and is in
general a non-local and anisotropic response function. In practice, determining
the effective dielectric function is a difficult problem, that can only be
solved under some (sometimes severe)
approximations~\cite{LakhtakiaBook,TsangBook}. In the present study, we do not
need to refer to a specific model. Our arguments rely only on the existence of
the effective medium for the propagation of the averaged field. The latter is a
direct consequence of the Dyson equation.

Under the following hypotheses, $({\it i})$ the random medium is isotropic on
average, and $({\it ii})$ only fields variations on scales larger than the size
of the scatterers and the correlation distance between scatterers are accounted
for, the dielectric function becomes a local and isotropic (scalar)
function~\cite{ShengBook,TsangBook}, i.e.
$\boldsymbol{\epsilon}^\perp_{\mathrm{eff}}(\kg) =
\boldsymbol{\epsilon}^\parallel_{\mathrm{eff}}(\kg) = \epseff \, \Ig$. We assume
that these conditions are satisfied in the following. The averaged Green
function in direct space now reads~:
\begin{equation}
\langle \Gg(\Rg) \rangle = PV \left [ \Ig+\frac{1}{\keff^2}\nabla\nabla \right ] \, \frac{\exp(i\keff R)}{4\pi R} - \frac{\Ig}{3 \keff^2} \delta(\Rg)
\label{eq:G}
\end{equation}
where $\keff=\neff \, k_0$, with $\neff=\sqrt{\epseff}$ the effective (complex)
refractive index of the random medium \cite{Bullough}.

%KRAMERS-KRONIG RELATIONS

From the principle of causality, one can derive the Kramers-Kronig (K-K) relations that connects the real and imaginary parts of the susceptibility of any linear material. Regarding the optical response, the K-K relations are usually written in terms of the dielectric function~\cite{Landau}. It can be shown that the refractive index in passive materials~\cite{comment} is also a quantity that satisfies the K-K relations (this amounts to showing that the refractive index is an analytic function in the upper half-plane of the complex frequency plane)~\cite{Toll56}. In a homogeneous medium, the imaginary part of the refractive index is associated with absorption (or reflection in the particular case of a non-absorbing electron gas below the plasma frequency). In a scattering medium and below the homogenization threshold, even in the absence of absorption, the effective index has a non-vanishing imaginary part that corresponds to extinction of the averaged field by scattering. The lost energy is redistributed in the field fluctuations, whose averaged square modulus is the diffuse intensity. In terms of the effective index, the K-K relations read~:
\begin{eqnarray}
\Real \neff(\omega) &=& 1 + \frac{2}{\pi} PV \int_0^\infty \frac{\omega^\prime \, \Imag \neff(\omega^\prime)}{\omega^{\prime 2}-\omega^2} \, \dd\omega^\prime
\label{eq:KK1} \\
\Imag \neff(\omega) &=& -\frac{2 \, \omega}{\pi} PV \int_0^\infty \frac{ \Real \neff(\omega^\prime)-1}{\omega^{\prime 2}-\omega^2} \, \dd\omega^\prime \ .
\label{eq:KK2}
\end{eqnarray}
As we will see, the K-K relations lead to interesting relationships between the spectra of the averaged DOS and the extinction MFP. Moreover, we will show that sum rules can be deduced under very general conditions.

%DOS AND EXTINCTION MFP

In an inhomogeneous medium, the LDOS at point $\rg$ is given by
$\rho(\omega,\rg) = 2 \omega/(\pi c^2) \, \Imag \mathrm{Tr} \, \Gg(\rg,\rg)$,
where $\mathrm{Tr}$ denotes the trace of a tensor~\cite{Lagen96,Lagen07}. The
LDOS counts the number of radiation states in the frequency range $[ \omega,
\omega+\dd\omega]$, weighted by their contribution at point $\rg$. Note that in
the near field, in the presence of evanescent states (due, e.g., to
electromagnetic surface modes), the expression of the LDOS needs to be
modified~\cite{RC03}. This subtlety is particularly important for the
calculation of thermal radiation energy densities~\cite{RC05}. In a macroscopic
random scattering medium with statistical translational invariance, i.e. far
from any surface boundary, the averaged LDOS does not depend on position and
equals the averaged DOS~:
\begin{equation}
\rho(\omega) = \frac{2 \omega}{\pi c^2} \, \Imag \mathrm{Tr} \langle \Gg(\Rg=0) \rangle \, .
\label{eq:defDOS}
\end{equation}
This quantity is the density of states that we consider in the present work. In
the particular case of a vacuum, the DOS $\rho_0(\omega)$ is deduced from
$\Gg_0$ using the identity $\lim_{R\to 0} \Imag \Gg_0(\Rg) = k_0/(6\pi) \,
\Ig$~\cite{RC06}. One recovers the well-known result $\rho_0(\omega) =
\omega^2/(\pi^2 \, c^3)$. From Eqs.~(\ref{eq:G}) and (\ref{eq:defDOS}), we
obtain the averaged DOS in the random medium following the same approach, using
the identity $\lim_{R\to 0} \Imag \langle \Gg(\Rg) \rangle= \keff/(6\pi) \,
\Ig$, which yields
%\begin{equation}
$\rho(\omega) =  \rho_0(\omega) \, \Real \neff(\omega)$.
%\label{eq:DOS}
%\end{equation}
The {averaged DOS is given by the real part of the effective refractive index.
We point out that this result has been obtained under the conditions of
existence of an isotropic and local effective refractive index. This means that
the effects of microscopic length scales on the order of the correlation length
or the size of the scatterers have been disregarded. In particular, this
expression coincides with the LDOS used to describe macroscopically the
spontaneous decay of dipole emitters in homogeneous dielectrics. It is known
that in dense materials local field corrections have to be incorporated in
order to account for interactions on microscopic length
scales~\cite{LocalFields}. In the following, we shall refer to the DOS given
above as the {\it averaged macroscopic DOS}.

The second important quantity in our discussion is the extinction MFP $\lext$,
that describes the attenuation of the averaged field by scattering and
absorption~\cite{ShengBook}. More precisely, $\lext$ is defined as the decay
length of the intensity of the averaged field. Therefore the extinction MFP is
given by the imaginary part of the effective refractive
index~\cite{TsangBook,Bullough}~:
%\begin{equation}
$\lext(\omega)= c/[2\,  \omega \, \Imag \neff(\omega)]$.
%\label{eq:MFP}
%\end{equation}

%DISPERSION RELATIONS AND SUM RULES

The principle of causality implies a close connection between the averaged macroscopic DOS and the extinction MFP. Indeed, by a direct application of the K-K relations Eqs.~(\ref{eq:KK1}) and (\ref{eq:KK2}), and using the expressions of $\rho(\omega)$ and $\lext(\omega)$ given above, we obtain~:
\begin{equation}
\frac{\rho(\omega)}{\rho_0(\omega)}= 1 + \frac{c}{\pi} PV \int_0^\infty \frac{[\lext(\omega^\prime)]^{-1}}{\omega^{\prime 2}-\omega^2} \, \dd\omega^\prime
\label{eq:KKDOS}
\end{equation}
\begin{equation}
\frac{1}{\lext(\omega)}= -4\pi\,  \omega^2 \, c^2 PV \int_0^\infty \frac{ \rho(\omega^\prime)-\rho_0(\omega^\prime)}{\omega^{\prime 2} \, (\omega^{\prime 2}-\omega^2)} \, \dd\omega^\prime \ .
\label{eq:KKMFP}
\end{equation}
These dispersion relations are the first result of this Letter. They are valid under very general conditions~: passive and causal medium, and assumption of an isotropic and local effective medium for the description of the averaged field. They demonstrate that the averaged macroscopic DOS and the extinction MFP are not independent. Equation (\ref{eq:KKDOS})  shows that from the spectrum of the MFP, one can deduce the averaged macroscopic DOS (and {\it vice versa} using Eq.~(\ref{eq:KKMFP})). In practice, this means that from an extinction spectrum (a natural measurement in spectroscopy), one could deduce the spectrum of the averaged macroscopic DOS.

From the K-K relations, sum rules for the dielectric constant and the refractive index can be obtained~\cite{Nussen72}. In particular, it is well established that the refractive index of any passive and causal medium satisfies $\int_0^\infty \omega \, \Imag \neff(\omega) \, [\Real \neff(\omega) -1] \, \dd\omega = 0$ and $\int_0^\infty [\Real \neff(\omega) -1] \, \dd\omega = 0$~\cite{Nussen72}. Beyond the principle of causality, the derivation of these sum rules relies on the assumption of a material behaving as a free-electron system in the high frequency limit~:
$\epseff(\omega) \sim 1- \omega_p^2/\omega^2$ when $\omega \to \infty$, where $\omega_p$ is an effective plasma frequency. Note that this hypothesis is not too restrictive, this high frequency behavior being expected as soon as the frequency is much larger than the resonant frequencies of the effective medium. The sum rules for the effective refractive index can be translated into new sum rules involving the averaged macroscopic DOS and the extinction MFP. In particular, we obtain~:
\begin{equation}
\int_0^\infty \frac{\rho(\omega)-\rho_0(\omega)}{\omega^2 \, \lext(\omega)} \, \dd\omega = 0 \ .
\label{eq:sumDOSMFP}
\end{equation}
This relation is the main result of this Letter. It demonstrates that the spectra of the averaged macroscopic DOS and of the extinction MFP are intimately connected, and constrained by a simple sum rule. The simplicity and the generality of this relation are striking. Let us remind the three conditions of validity~: (1)~the medium has to be passive and causal, (2)~the effective medium is described by an isotropic and causal dielectric function, (3)~the medium behaves as a free electron gas in the high frequency limit. The second sum rule for the refractive index leads to a relation involving the averaged macroscopic DOS only~: $\int_0^\infty  [\rho(\omega)-\rho_0(\omega)]/\omega^2 \, \dd\omega = 0$. This second sum rule was established previously in the context of spontaneous emission in dielectric media~\cite{Loudon96}. Regarding wave propagation in random media, it establishes a constraint on the potential modifications of the averaged macroscopic DOS. In particular, it shows that $\rho(\omega)$ is necessarily lower than the DOS in free space $\rho_0(\omega)$ in a spectral range, and greater than $\rho_0(\omega)$ in another spectral range (the numerator $\rho(\omega)-\rho_0(\omega)$ has to change sign somewhere for the integral to vanish). This behavior is also dictated by Eq.~(\ref{eq:sumDOSMFP}) because $\lext(\omega)>0$ in a passive medium. In the following, we illustrate the general behavior induced by relations (\ref{eq:KKDOS}-\ref{eq:sumDOSMFP}) in a particular case.

%NUMERICAL EXAMPLE

Let us consider a random scattering medium with an effective dielectric function exhibiting a resonance at a particular frequency $\omega_0$. The resonance can be induced by an internal resonance of the scatterers, or of purely geometric origin (or both). We choose a Lorentz model of the form~:
\begin{equation}
\epseff(\omega) = 1 + \frac{{\cal F} \, \omega_0^2}{\omega_0^2-\omega^2-i \, \omega \, \Gamma}
\label{eq:epseff}
\end{equation}
where the parameter ${\cal F}$ is an effective oscillator strength and $\Gamma$ is the linewidth. From this expression, the effective index $\neff=\sqrt{\epseff}$ is readily obtained numerically, as well as the spectra of the averaged DOS $\rho(\omega)$ and the extinction MFP $\lext(\omega)$. As noted in Ref.~\cite{Lagen96}, a regime of strong scattering is identified when $\Imag \neff(\omega) > \Real \neff(\omega)$, which corresponds to $k  \lext < 1/2$, where $k=\Real \keff$ is the wavenumber in the medium.  In this regime, the effective medium satisfies $\Real \epseff(\omega)<0$ and the averaged field is strongly damped (the effective medium has a metallic character). In the regime $k  \lext > 1/2$, one has $\Real \epseff(\omega) >0$ and the effective medium has a dielectric character. The transition between these two regimes is driven by the parameter ${\cal P} = {\cal F} \, \omega_0/\Gamma$. For ${\cal P} <2$, one has $\Real \epseff(\omega) >0$ at all frequencies. For ${\cal P} >2$, there is a frequency range for which $\Real \epseff(\omega) <0$, or equivalently $k  \lext < 1/2$.

We show in Fig.~\ref{fig:P1} the spectra of the averaged macroscopic DOS (left) and of the extinction MFP (right), in the case of an effective medium of dielectric character (${\cal P}=1$). We observe that due to the resonance, the extinction MFP exhibits a minimum. Due to the K-K relationship, the averaged macroscopic DOS shows a typical dispersion behavior. This illustrate the fact that the DOS has necessarily a complicated structure around a minimum of the extinction MFP.
\begin{figure}[h]
\psfrag{d/G}{\small{$\delta/\Gamma$}}
\includegraphics[width=8cm]{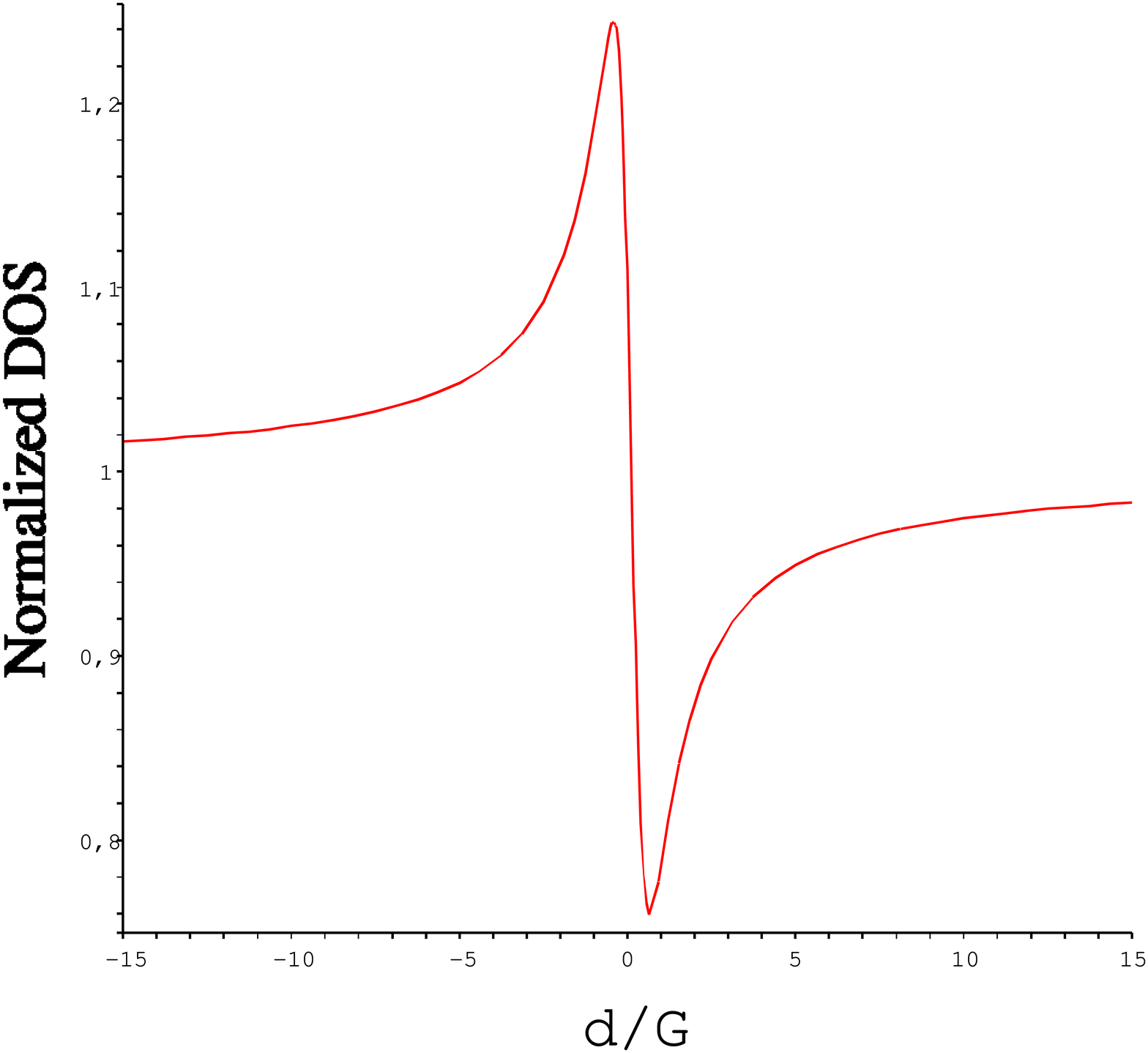}
\includegraphics[width=8cm]{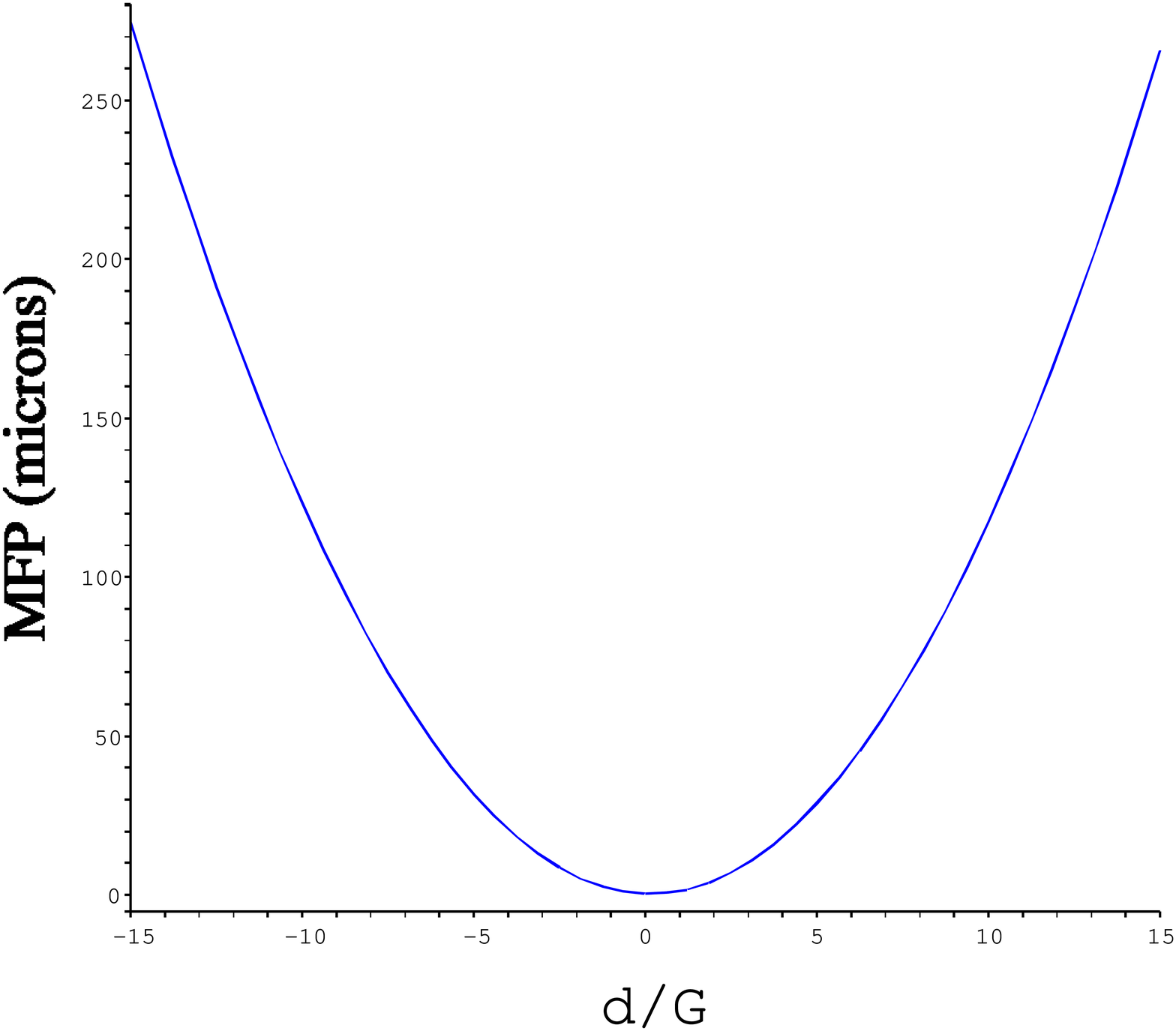}
\caption{ \label{fig:P1} Normalized DOS $\rho(\omega)/\rho_0(\omega)$ (left) and extinction MFP $\lext(\omega)$ in $\mu$m (right) versus the normalized detuning from resonance $\delta/\Gamma=(\omega-\omega_0)/\Gamma$. The effective dielectric function $\epseff(\omega)$ is given by Eq.~(\ref{eq:epseff}), with $\omega_0=10^{15}$ Hz, $\Gamma = 10^9$ Hz  and ${\cal P}=1$.}
\end{figure}
In Fig.~\ref{fig:P20}, we plot the same quantities for a medium with a strong metallic character (${\cal P}=20$). The extinction MFP exhibits a region with a low value, corresponding to the spectral region for which $\Real \epseff(\omega)<0$ or $k  \lext < 1/2$. In the same spectral range, the DOS is minimum, showing the existence of a pseudo-gap. Moreover, close to the lower pseudo-gap band edge, the DOS exhibits a strong oscillation that is a feature of the underlying dispersion relation.
\begin{figure}[h]
\psfrag{d/G}{\small{$\delta/\Gamma$}}
\includegraphics[width=8cm]{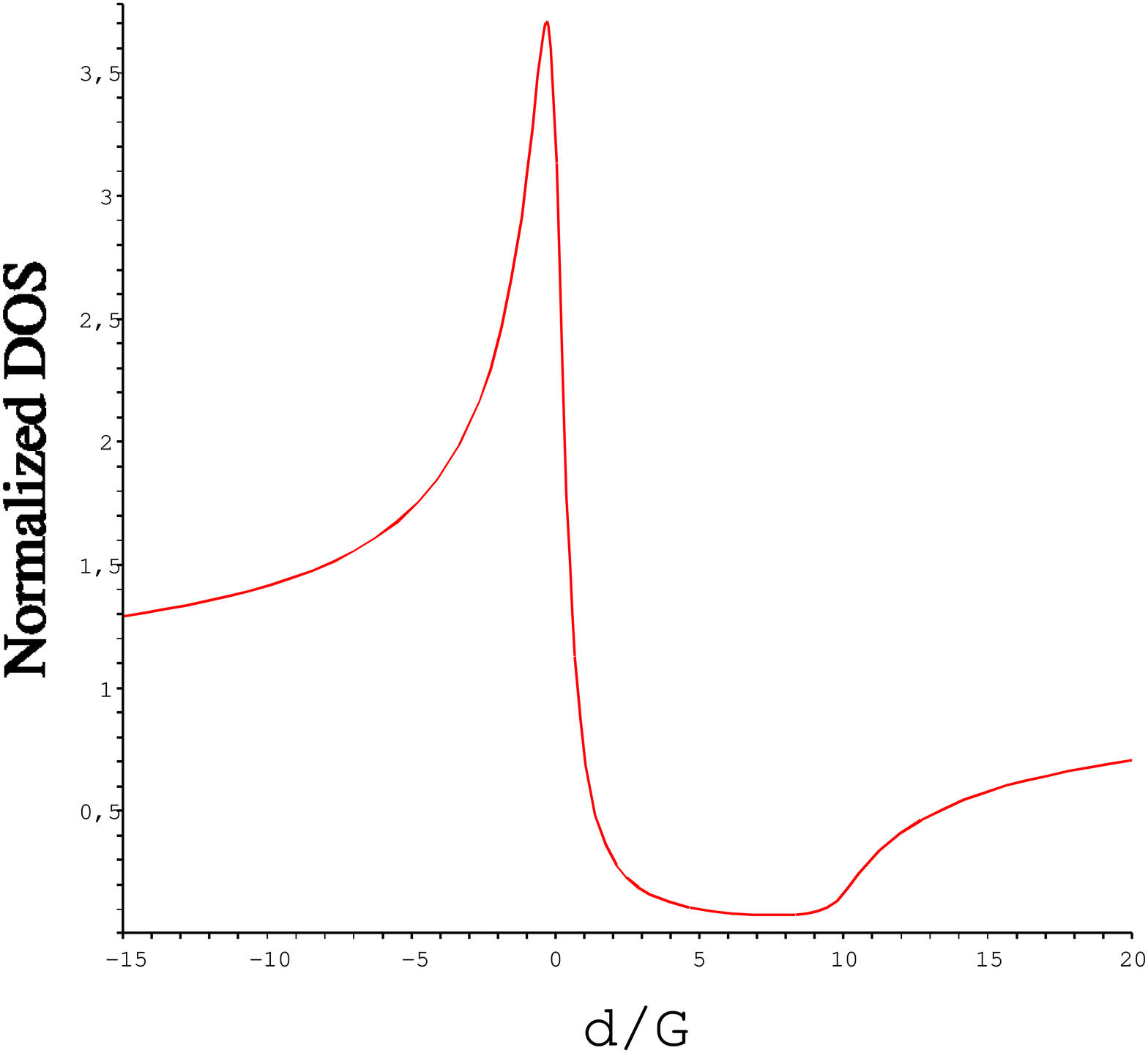}
\includegraphics[width=8cm]{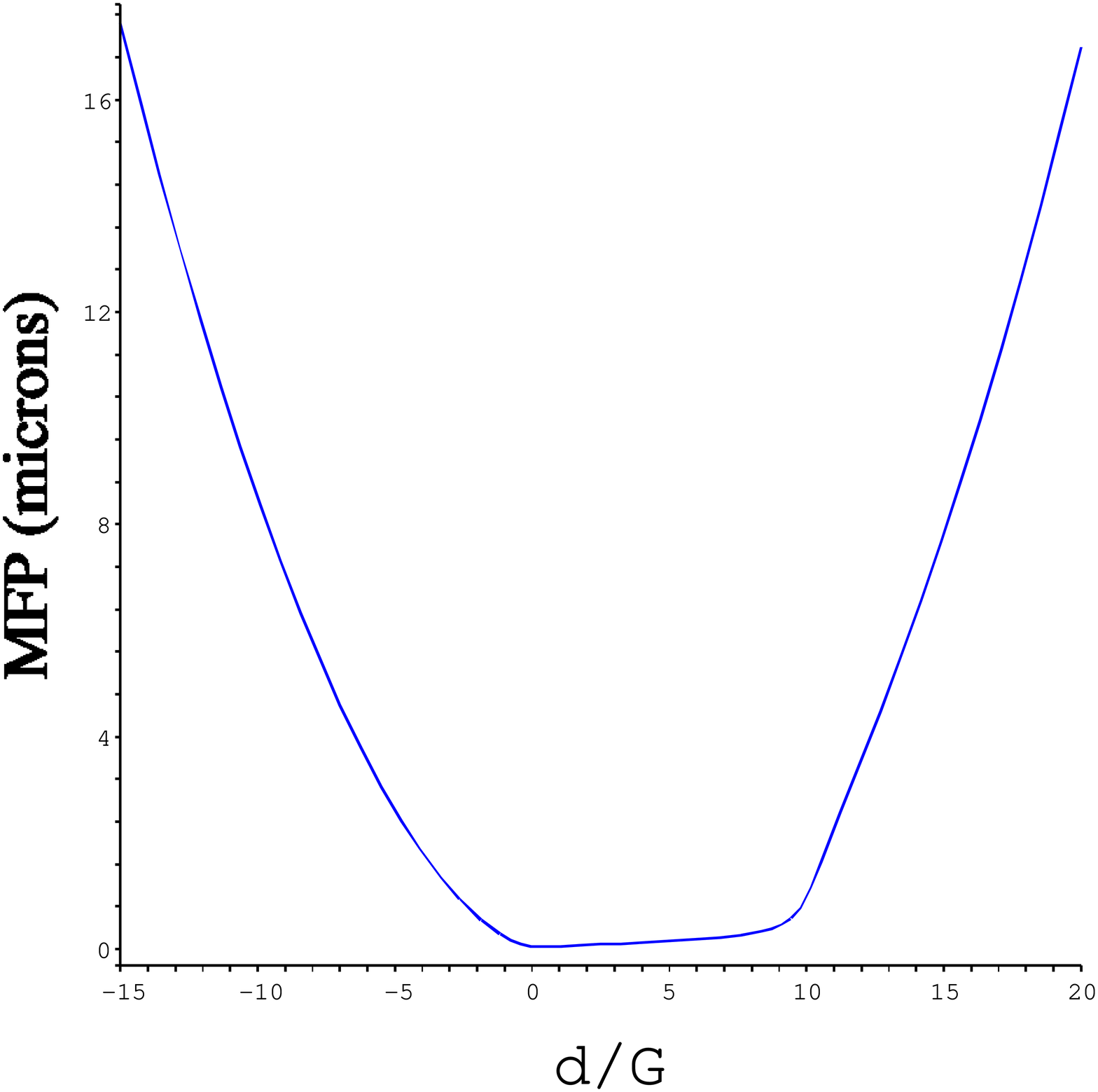}
\caption{ \label{fig:P20} Same as Fig.~1 with ${\cal P}=20$.}
\end{figure}

Finally, we note that in the case where the effective dielectric constant is known analytically, one has $\Imag \epseff = 2 \, \Imag \neff \, \Real \neff$ at each frequency. In terms of averaged DOS and MFP, this leads to the explicit relation $\Imag \epseff(\omega) = c/[\omega \lext(\omega)] \, \rho(\omega)/\rho_0(\omega)$. This relation is useful when an analytical model of $\epseff$ is available, a situation that occurs only in a few specific cases.

%CONCLUSION

In summary, from the principle of causality, we have established dispersion relations and a frequency sum rule that connect the averaged macroscopic DOS and the extinction MFP in a random scattering medium. These relations constrain the spectral variations of both quantities, and sustain general features of their spectral behavior close to a resonance of the effective medium. These results should be helpful in the analysis of measurements of wave transport through complex systems and in the design of randomly or periodically structured materials with specific transport properties.

We acknowledge helpful discussions with S. Albaladejo, L.S. Froufe-P\'erez and
F. Scheffold. This work was supported by the EU Integrated Project
\textit{Molecular Imaging} under contract LSHG-CT-2003-503259, the Spanish
Consolider \textit{NanoLight} (Ref. CSD2007-00046) and by the French ANR
project CAROL BLAN06-3-134124.

\vspace{-0.3cm}

%%%%%%%%%%%%%%

\end{document}